\documentclass[fleqn, 11pt]{wlscirep}
\pdfoutput=1

\usepackage[utf8]{inputenc}
\usepackage[T1]{fontenc}
\usepackage{lipsum,gensymb,multirow,lineno,float,amsmath,flushend,lmodern,xr,xcolor,color,soul,graphicx,subfigure,esvect,miller,breakurl,microtype,siunitx}
\usepackage[export]{adjustbox}
\usepackage[version=4]{mhchem}
\usepackage[figuresright]{rotating}
\usepackage{hyperref}
\hypersetup{
    breaklinks=true,
    colorlinks=true,
    linkcolor=blue,
    filecolor=magenta,      
    urlcolor=cyan,
    pdftitle={Revealing 3D Strain and Carbide Architectures in Additively Manufactured Ni Superalloys},
    pdfpagemode=FullScreen,
    }

\DeclareMathAccent{\wtilde}{\mathord}{largesymbols}{"65}

\DeclareSIUnit{\pixel}{pixels}
\DeclareSIUnit{\voxel}{voxels}
\DeclareSIUnit{\angstrom}{\AA}
\DeclareSIUnit\weightpercent{\text{wt.}\%}

\DeclareUnicodeCharacter{03B1}{$\alpha$}
\DeclareUnicodeCharacter{03B3}{$\gamma$}
\DeclareUnicodeCharacter{025B}{$\varepsilon$}
\DeclareUnicodeCharacter{03B5}{$\varepsilon$}

\usepackage{pdfpages} 
\usepackage{pgffor} 

\makeatletter
\AtBeginDocument{\let\LS@rot\@undefined}
\makeatother

\def\supplementfilename{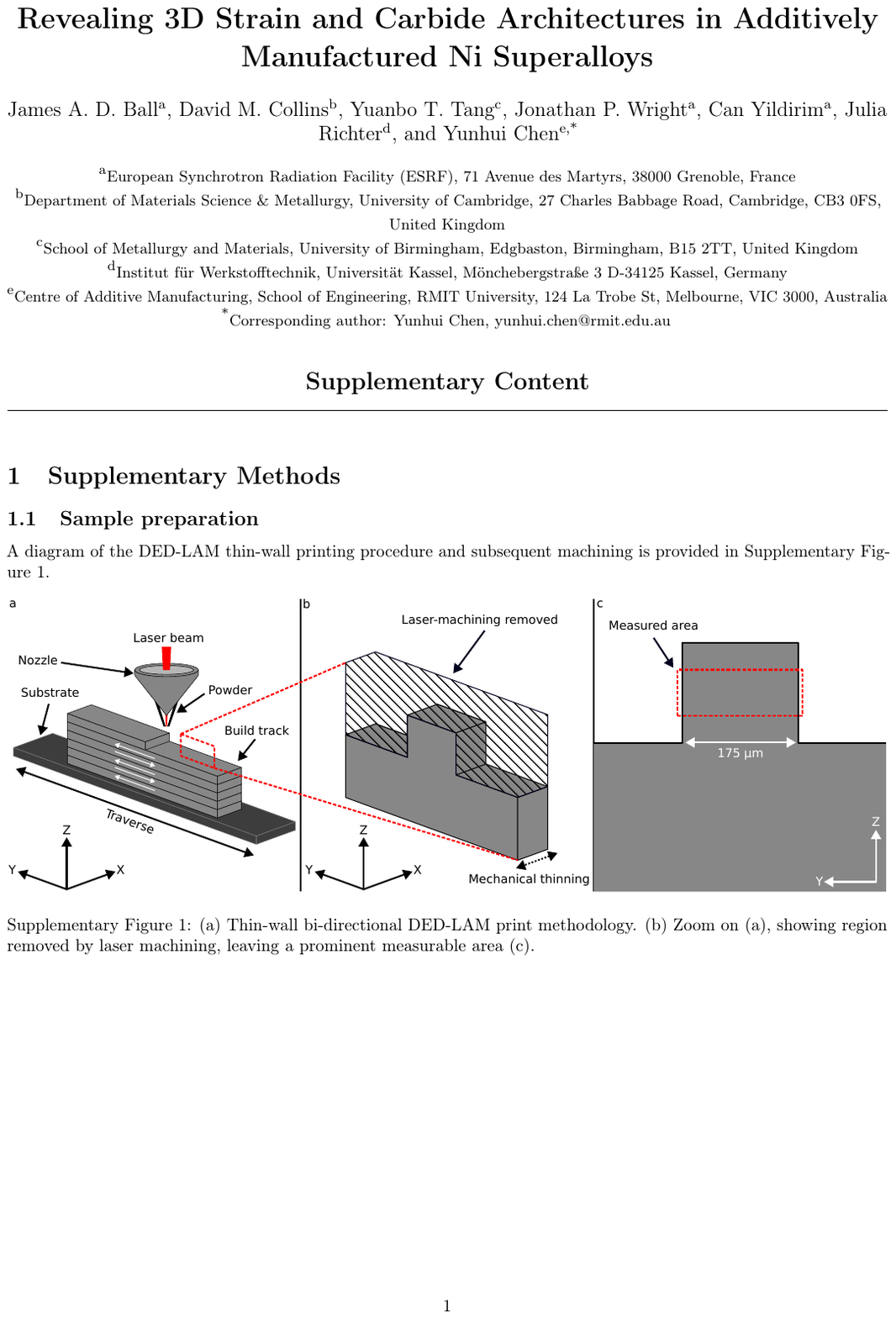}
\pdfximage{\supplementfilename}
\def\numbersupplementpages{\the\pdflastximagepages}
\newif\ifarXiv
\arXivtrue 

\title{Revealing 3D Strain and Carbide Architectures in Additively Manufactured Ni Superalloys}

\author[1]{James A. D. Ball}
\author[2]{David M. Collins}
\author[3]{Yuanbo T. Tang}
\author[1]{Jonathan P. Wright}
\author[1]{Can Yildirim}
\author[4]{Julia Richter}
\author[5,*]{Yunhui Chen}

\affil[1]{European Synchrotron Radiation Facility (ESRF), 71 Avenue des Martyrs, 38000 Grenoble, France}
\affil[2]{Department of Materials Science \& Metallurgy, University of Cambridge, 27 Charles Babbage Road, Cambridge, CB3~0FS, United Kingdom}
\affil[3]{School of Metallurgy and Materials, University of Birmingham, Edgbaston, Birmingham, B15~2TT, United~Kingdom}
\affil[4]{Institut für Werkstofftechnik, Universität Kassel, Mönchebergstraße 3 D-34125 Kassel, Germany}
\affil[5]{Centre for Additive Manufacturing, School of Engineering, RMIT University, 124 La Trobe St, Melbourne, VIC 3000, Australia}
\affil[*]{yunhui.chen@rmit.edu.au}
        

\begin{abstract}
Fast directional solidification during Laser Additive Manufacturing (LAM) produces a complex microstructure in nickel-based superalloys, comprising columnar grains with cellular sub-grain structures and carbides. 
Using non-destructive Scanning 3D X-ray Diffraction (S3DXRD), we reveal spatially complex orientation and intergranular strain relationships that couple strongly to processing-induced cellular sub-grain networks and a primary cubic metal carbide (MC) phase. 
We have examined 3D orientation and elastic strain tensor fields across 82 $\gamma$ grains together with the spatial distribution of over 37,000 MC carbides in an ABD-900AM alloy sample manufactured by the Directed Energy Deposition (DED) LAM process.
Carbides are spatially associated with the cellular sub-grain network with a weak but present orientation relationship with their parent $\gamma$ grains.
The MC carbides, known to be Ti, Ta and Nb rich, form in regions of high solute segregation, resulting in a significant volumetric lattice parameter patterning in the associated $\gamma$ phase regions.
These chemically distinct solute-rich regions possess a higher associated elastic modulus compared to intercellular regions and determine the local residual stress patterning.
These results provide the first non-destructive 3D study of the relationship between rapid solidification-induced segregation, deformation heterogeneity and carbide architectures in an additively manufactured Ni-based superalloy.
The insights provide crucial detail to rationalise LAM process parameter optimisation and the coupled spatially governed structural performance.
\end{abstract}

\begin{document}
\flushbottom
\maketitle
\newpage

\section*{Introduction}
Laser Additive Manufacturing (LAM) methods such as Laser Powder Bed Fusion (L-PBF) and Directed Energy Deposition (DED-LAM) enable the rapid production of near-net-shape components with complex geometries, offering substantial reductions in material waste and buy-to-fly ratios compared to conventional subtractive methods \cite{mostafaei_additive_2023, bridges_metallurgical_2020, debroy_additive_2018, panwisawas_metal_2020}.
These advantages make LAM attractive for high-value aerospace applications, particularly for Ni-based superalloy components requiring the most demanding high-temperature performance \cite{english_overview_2010, pollock_alloy_2016}.
Beyond reducing cost and waste, DED-LAM in particular enables the repair of critical components such as turbine blades \cite{saboori_application_2019}.
DED-LAM provides faster build rates, larger melt passes than L-PBF and the flexibility for localised or site-specific builds, making it particularly suitable for large components and repair strategies.

Despite this promise, LAM of Ni-based superalloys remains challenging.
Rapid, highly directional solidification that is inherent to the processes leads to anisotropic mechanical properties due to strong crystallographic texture \cite{tang_effect_2020, chen_high-resolution_2023, deng_microstructure_2018}, interdendritic cracking \cite{tang_alloys-by-design_2021}, and substantial residual strains \cite{chen_high-resolution_2023, chen_correlative_2021,  cottam_characterization_2014, yin_laser_2021, yu_thin-wall_2020}.
At smaller length scales, cellular sub-grain structures form; a feature that has been observed across several DED-LAM alloy systems including stainless steels \cite{kong_about_2021, wang_additive-manufacturing-induced_2025, chen_strengthening_2022, tognan_3d_2026}, \ce{Al}-based alloys \cite{prashanth_microstructure_2014, thijs_fine-structured_2013}, \ce{CoCrMo} alloys \cite{chen_grain_2017, qian_defects-tolerant_2015, hedberg_vitro_2014}, and \ce{Ni}-based superalloys \cite{chen_high-resolution_2023, tang_alloys-by-design_2021, chen_dendritic_2016, liu_strengthening_2023}.
Honeycomb-like sub-grain networks emerge due to constitutional undercooling conditions when the preferred growth direction (\hkl<100> in cubic systems) aligns with the maximum thermal gradient \cite{kong_about_2021, chen_grain_2017, kou_weld_2002}.
The result is a highly inhomogeneous sub-grain microstructure -- the cell boundaries, which solidify last, are solute-rich with high densities of dislocations and primary or secondary phase particles, while the cell cores are softer and solute lean \cite{chen_correlative_2021, kong_about_2021, liu_strengthening_2023, kou_weld_2002, tang_multi-length-scale_2023, liu_characterization_2020}.
Cellular structures have attractive characteristics -- they improve the strength of as-built components via hetero deformation-induced strengthening, which includes several underlying mechanisms: dislocation density modification \cite{kong_about_2021,  tang_multi-length-scale_2023, liu_dislocation_2018,  zhang_microstructure_2024, wang_additively_2018}, solute segregation \cite{liu_strengthening_2023, wang_additively_2018, cahn_hardening_1963}, and precipitation strengthening \cite{zhang_strengthening_2021, voisin_new_2021, tian_excellent_2025}.
In DED-LAM-produced Ni alloys, the crystal structure and stoichiometry of inhomogeneous cell-boundary phases depend on the bulk composition.
However, when elemental partitioning promotes the formation of phases such as Laves, $\delta$ and MC, mechanical properties, including impact hardness, fatigue life, and fracture toughness, are significantly compromised \cite{tian_rationalization_2014, gribbin_role_2019, nezhadfar_fatigue_2020, yu_influence_2020, sui_influence_2017, yu_microstructure_2021}.
The spatial distribution of precipitates is also an important factor; as an example, micron-sized MC carbides in DED-LAM IN718 have recently been shown to be a primary source of fracture \cite{ahmad_revealing_2022}.
They have also been shown to modify the corrosion pathways due to the heterogeneous solute concentrations \cite{liu_corrosion_2024, delvecchio_metastable_2024}.

Given the importance of the cellular structure in LAM materials, there have been several recent studies that have sought to obtain three-dimensional morphologies, which have been inferred through two-dimensional imaging and Electron Back-Scatter Diffraction (EBSD) methods \cite{wang_additive-manufacturing-induced_2025, lia_thermal_2018, wang_cell_2025}, as well as 3D modelling \cite{ren_solute_2023, wei_three-dimensional_2019}.
Investigations of precipitate phase distributions and crystallographic orientations, though mature and varied, have been almost entirely limited to two dimensions, primarily due to their very small size compared with the cellular length scales in play \cite{tang_multi-length-scale_2023, tian_excellent_2025}.
Exceptionally, recent impressive studies using TriBeam serial sectioning and a combination of EBSD and Scanning Electron Microscopy (SEM) methods have enabled direct large-volume three-dimensional measurements of grain orientations and geometrically necessary dislocation (GND) densities in L-PBF-produced Ni-based superalloys \cite{polonsky_defects_2018, polonsky_3d_2020, wittenzellner_microstructural_2020, lamb_analysis_2024}.
While these studies were also able to capture precipitate phase distributions via additional serial sectioning and imaging \cite{polonsky_defects_2018}, their orientations were not described, and particles decorating individual L-PBF cells were not detected, likely due to a highly refined cellular diameter that was below the detection limit (typically \qty{< 1}{\micro\metre} \cite{zhang_microstructure_2024}).
It remains unclear whether phases such as carbides are coherently oriented with the $\gamma$ matrix, which would modify their dislocation pinning ability \cite{zhang_strengthening_2021, tian_excellent_2025, tian_rationalization_2014}, and potentially how their distribution couples to local strain fields.
The EBSD scans were also limited in spatial resolution to $\qty{1.5}{\micro\metre}$, and although GND densities could be computed via per-voxel orientation measurements, elastic strain fields were not resolved.
Furthermore, mechanical sectioning can alter the strain fields, rendering the results unreliable.
Recently, non-destructive methods such as Dark Field X-ray Microscopy (DFXM) have provided valuable insights into 3D elastic strain fields and cellular morphologies in LAM-produced Ni-based superalloys \cite{chen_high-resolution_2023}.
However, measurements are constrained to high-resolution scans of individual grains -- only one component of the strain tensor is accessible, making the technique less sensitive to the presence of small particles or precipitates.

A comprehensive 3D dataset that captures the cellular morphology, strain fields from solute partitioning, and precipitate or particle phase distributions with orientations is essential for predictive modelling, alloy development, and resource-efficient manufacturing.
Here, we present the first scanning 3D X-ray diffraction (S3DXRD) study of an LAM-processed (DED-LAM) Ni-based superalloy -- ABD-900AM -- designed for high printability \cite{tang_alloys-by-design_2021, crudden_nickel-based_2019, reed_alloys-by-design_2009}.
The dataset captures full 3D orientation and strain tensor fields across hundreds of $\gamma$-phase (A1 crystal structure, using Strukturbericht designation) grains together with the indexed positions, orientations and strain states of \num{\sim 37000} MC carbides.
This comprehensive volume-resolved dataset enables direct examination of strain and orientation dependencies of both the $\gamma$ matrix and carbides from a sample comprising multiple build tracks.
This offers a new lens through which to understand microstructure–property relationships in LAM of Ni-based superalloys.

\section*{Results}

\begin{figure}[h]
    \centering
    \includegraphics[width=\textwidth]{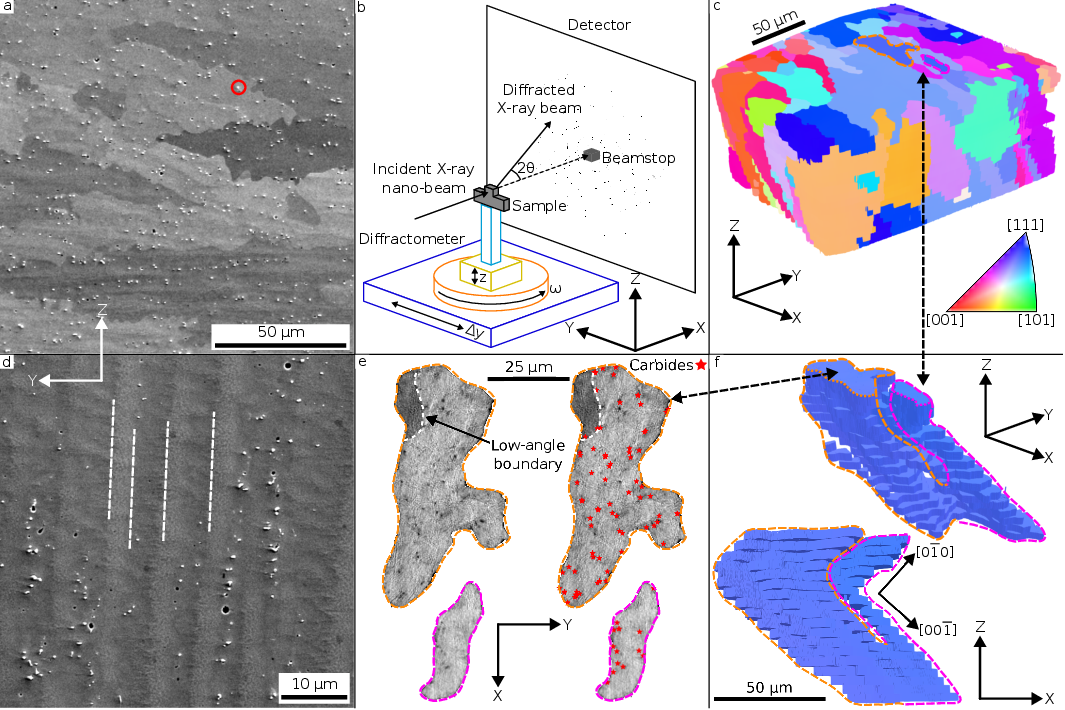}
    \caption{\textbf{Correlative SEM and S3DXRD of Ni-based superalloy microstructure comprising a $\gamma$ matrix and primary MC carbides.} (a, d) Backscatter electron micrographs of a sectioned sample with build direction parallel to the labelled $z$-axis, at low (a) and high (d) magnifications. Annotated features include an example MC carbide circled in red in (a) and sub-grain cellular structures highlighted in white in (d). (b) Sample mounted on Nanoscope diffractometer (ID11, ESRF) showing stage names and the laboratory coordinate system. (c) Reconstructed S3DXRD volume, $\gamma$ phase, in IPF-$Z$ colouring. (f) Single $\gamma$ grain with two dendrites (orange and magenta) isolated from (c), two perspectives. (e) Top $Z$-layer of $\gamma$ grain from (f), diffracted intensity colouring (left) with positions of MC overlaid in red (right).}
    \label{fig:1}
\end{figure}

\subsubsection*{Microstructures}

S3DXRD was employed to capture three-dimensional localised crystallographic orientation, strain fields, and phase information. A schematic of the experimental configuration is shown in Figure~\ref{fig:1}b, with details of the acquisition process provided in the Method section.
Each S3DXRD layer was acquired via a sequence of sample rotations -- the sample was rotated \qty{360}{\degree} in $\omega$ with a step size $\delta\omega = \qty{0.125}{\degree}$ while diffraction images were collected. In total, \num{18} layers were collected to give a measured volume of \qtyproduct{200 x 175 x 90}{\micro\metre\cubed} with a \qty{5}{\micro\metre} $Z$ (vertical) spacing between layers.
The sample was oriented such that the height of the sample (build direction, see Supplementary Figure 1) was approximately parallel to the laboratory $Z$ axis.

To identify the phases, histograms of the observed diffracted intensity vs $d^*$ (inverse $d$-spacing) across the full set of segmented diffraction peaks in all S3DXRD layers were generated, yielding a one-dimensional intensity profile; this is shown in Supplementary Figure 6.
Peak indexing confirmed the microstructure comprised $\gamma$ and cubic MC phases only. 
There is no evidence of the L1$_2$ structured $\gamma'$ precipitates, which would be present under thermodynamic equilibrium conditions \cite{tang_alloys-by-design_2021}. 
The Laves phase was also not observed.
Thus, the as-built DED-LAM Ni-based superalloy sample studied here has a non-equilibrium microstructure with a highly supersaturated $\gamma$ matrix.

The microstructure was spatially quantified by tomographic reconstruction of the S3DXRD measurements.
The sample contained \num{82} $\gamma$ grains and \num{37944} MC carbides, recorded across all $Z$ layers with a mean of \num{60} $\gamma$ grains per $Z$ layer.
A merged map of $\gamma$ grains is shown in Figure~\ref{fig:1}c, with grains coloured by their positions on the $Z$-axis Inverse Pole Figure (IPF-$Z$).
A highly columnar microstructure is observed, with the long axis inclined close to the $Z$ axis (the build direction).
The columnar feature of the grains was more evident when observing an isolated grain, as shown in Figure~\ref{fig:1}f.
Two \hkl<100> directions for the grain of interest are included, indicating a distinctive change in growth direction -- this is likely attributed to a switch in the thermal gradient directions during the build process, while the laser pass direction reverses on the bi-directional thin wall build.
In Figure~\ref{fig:1}e, a single layer from this grain is shown, coloured by the diffraction intensity propagated by the tomographic reconstruction.
Areas of low diffraction intensity (darker regions) within the $\gamma$ grain are clearly visible -- small spots of reduced intensity indicate the presence of an additional phase. 
By overlaying the centre-of-mass positions of the MC carbides associated with this grain in this $Z$-layer, they align well with the intensity gaps.
These S3DXRD observations corroborate the SEM measurements in Figure~\ref{fig:1}a~\&~Figure~\ref{fig:1}d, which confirms a complex microstructural morphology, exhibiting tortuous grain boundaries with a high density of carbides that are dispersed at both intragranular and intergranular locations.

\subsubsection*{Residual strain}

\begin{figure}[h!]
    \centering
    \includegraphics[width=\textwidth]{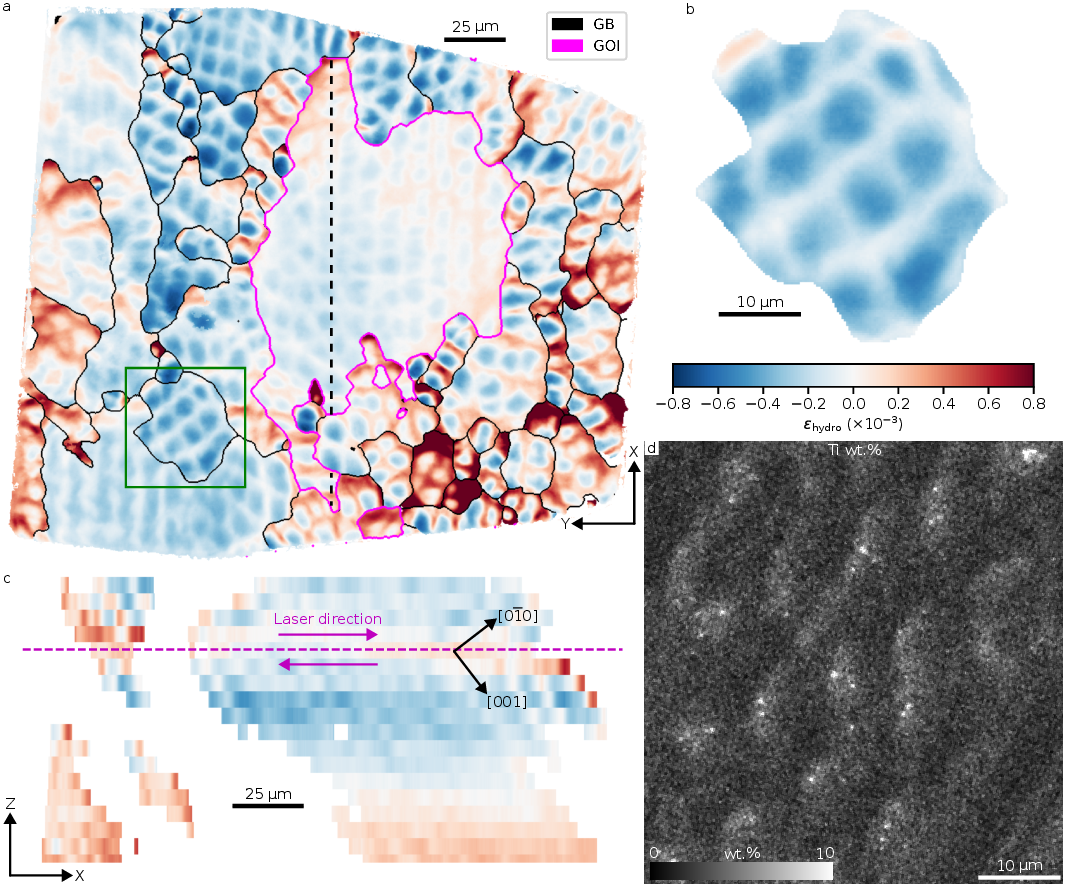}
    \caption{\textbf{Unit cell length fields of the $\gamma$ matrix reveal cellular sub-grain structures,} via hydrostatic strain magnitude. evident when viewed as (a) a single $Z$ layer, $\gamma$ phase, with the grain boundaries (GB) and a grain of interest (GOI) indicated. In (b), a magnified single grain is shown, and (c), the GOI viewed along the transverse to the vertical dashed line in (a). The inhomogeneity of the solutes is evident in (d) with an EDS map of the Ti concentration (wt.\%) as an example element, showing the concentration field matching the length scale of the hydrostatic strain modulation.}
    \label{fig:2}
\end{figure}

Strain refinement \cite{henningsson_microstructure_2024} was performed on the S3DXRD dataset.
By extracting the hydrostatic component of the strain tensors, the change in local unit cell length ($\varepsilon_{\rm hydro} = \frac{da_0}{a_0}$) can be directly investigated, as shown in Figures~\ref{fig:2}a, b and c. A clear sub-grain cellular morphology is observed in the hydrostatic strain distribution.
Interestingly, the cell boundaries have $\varepsilon_{\rm hydro} \approx 0$, indicating a lattice parameter very close to the size of the reference unit cell (\qty{3.589}{\angstrom}).
Cell cores, in contrast, display consistently reduced hydrostatic strains, indicating a substantial depletion of the solute elements in these regions, induced by the rapid non-equilibrium solidification during the build process.
These observations have been validated in the Energy Dispersive Spectroscopy (EDS) map (Figure~\ref{fig:2}d).
The Ti distribution map is presented -- area maps of other elements are shown in Supplementary Figure 7 with a line scan across a cell core in Supplementary Figure 8. Chemical segregation between cell boundaries and cell cores is clearly observed, which is highly correlated with the strain fields.

A variation of grain and cell size and shape was observed within the sample cross-section, hypothesised to be due to changes in cooling rates and deposition directions during the building process. 
These variations correlate with the strain magnitudes, corroborating other 3D polycrystalline studies that showed larger strains are generally associated with smaller grains (e.g. \cite{ball_revealing_2024}). 
Observations in this study are limited to just a few grains, so a generalised trend cannot be inferred. 
In our sample, where the hydrostatic strain field of a large central grain is plotted along the vertical build direction (Figure~\ref{fig:2}c), a clear boundary layer is observed, which is associated with an increased overall hydrostatic strain -- this could be explained by a deposition direction change during the bi-directional print that may have been captured within the measured volume.
The grain growth directions align well with the \hkl<001> preferential growth directions \cite{zhao_influence_2011} annotated in Figure~\ref{fig:2}c, supporting this hypothesis.

\subsubsection*{MC distribution and orientations}

\begin{figure}[h]
    \centering
    \includegraphics[width=\textwidth]{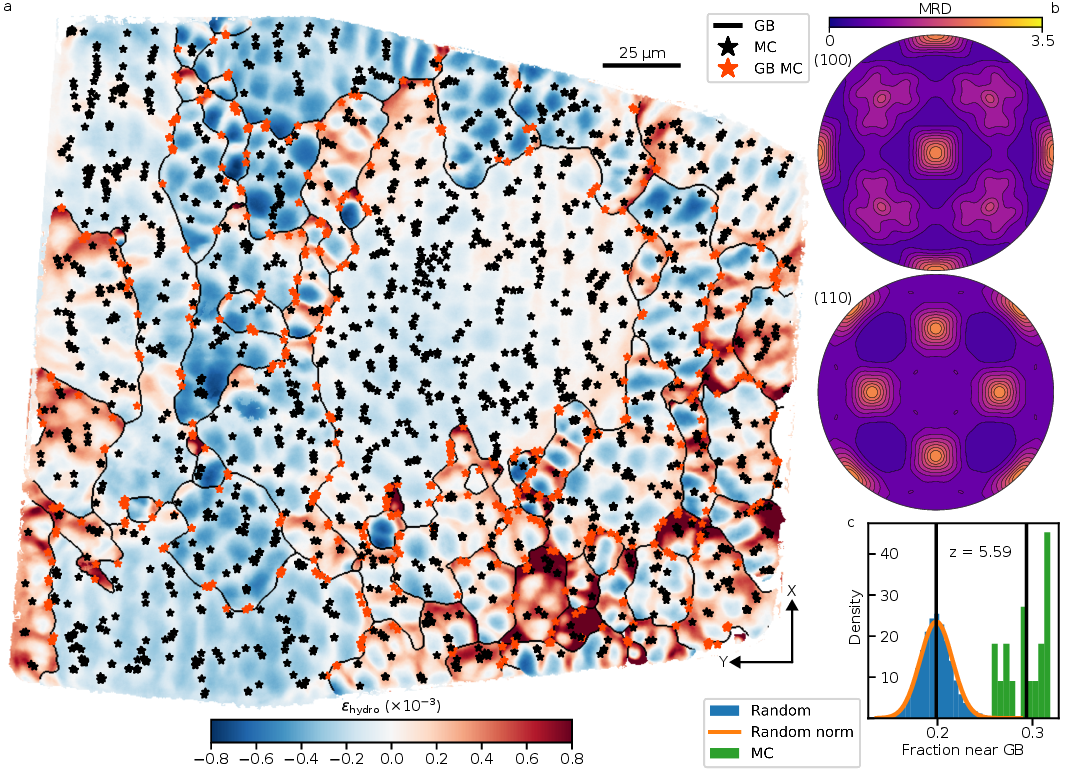}
    \caption{\textbf{Spatial and orientation distributions of MC carbides.} (a) Single $Z$ layer with hydrostatic strain colouring for the $\gamma$ phase with the grain boundaries (GB) highlighted. The MC carbide positions are overlaid in red (when near to a GB) and black (otherwise). (b) Pole figures of misorientation density functions for the MC carbides with respect to their parent $\gamma$ grains, coloured by multiples of random distribution (MRD). (c) Fraction of MC carbides (all layers) located within \qty{1}{\micro\metre} of grain boundaries (MC) versus the expected distribution for random points (Random). The $Z$-score indicates the number of standard deviations between the random and observed distributions.}
    \label{fig:3}
\end{figure}

In Figure~\ref{fig:3}a, carbide locations are denoted on the hydrostatic strain field.
Note that the markers indicate the centre of mass positions of the MC carbides only; their volumes cannot be reliably determined from the experimental data, due to their variable compositions. 
The carbides preferentially decorate the cell and grain boundaries, corresponding to the regions of increased hydrostatic strain and chemical segregation.
Some preferential grain boundary decoration is also present, as explored quantitatively in Figure~\ref{fig:3}c.
By computing the likelihood that a random position in the sample volume would touch a grain boundary, the ``Random'' distribution was generated.
When computing the observed spatial distribution of the MC carbides, a clear preference for grain boundary decoration is evident, at a significance of more than 5 standard deviations from the random distribution.

By fitting an orientation distribution function to the MC-$\gamma$ misorientations for each carbide, pole figures were generated to infer the orientation relationship (OR) between each MC carbide and its parent $\gamma$ grain, as shown in Figure~\ref{fig:3}b.
The misorientation pole figures indicate a cube-cube relationship, inferred from the 4-fold symmetry for each of the \hkl(100) \& \hkl(110) pole figures.
However, the peak Multiples of Random Distribution (MRD) value of $\approx3.5$ reveals that this is not a particularly strong orientation relationship and that a large portion of the carbides deviate from the cube-cube OR.
This is supported by the texture index $t$ of the misorientation distribution function $f(R)$, computed by the MTEX software \cite{hielscher_mtex_2007} via $t=\int_{{\mathrm{SO}}(3)}{f(R)^2}{{\rm d}R}$ (where $\mathrm{SO}(3)$ is the group of all 3D rotations), which was found to be $t=1.77$, indicating a weak misorientation texture.

\section*{Discussion}

This work presents the first direct three-dimensional assessment of the coupled strain fields, carbide spatial arrangement, and grain morphology in a DED-LAM Ni-based superalloy.
This example dataset represents a new paradigm for quantitative inputs used in process optimisation models that seeks to explain, for example, solidification-driven microstructural evolution of laser additively manufactured alloys. 
Using S3DXRD, these data further support our aim to understand the DED-LAM intragranular relationships between microstructure and spatially varying crystallography. 
These aspects, alongside a critique of the S3DXRD measurements, are next considered.

\subsubsection*{Solute segregation}

Features of the observed microstructure have been previously reported \cite{chen_high-resolution_2023}, comprising a dendritic microstructure with sub-grain cells, a distribution of MC carbides, and no $\gamma'$ precipitates (Supplementary Figure 1).
High thermal gradients arising from rapid cooling rates during DED-LAM determine the liquid to solid transformation rates, here favouring dendritic solidification, as shown in Figure~\ref{fig:1}.

Constitutional supercooling \cite{ren_solute_2023}, which is influenced by alloy composition, particularly solute content and partitioning behaviour, dictates the cell size and segregation patterning in our sample.
EDS analysis confirms Ti and Nb enrichment at cell boundaries ($\sim$+2.5\,wt.\%), with a concomitant depletion of Ni, Co and Cr. 
The reader is referred to Supplementary Figures 7--8 for EDS maps and line scans of multiple elements.
A similar result was observed for this alloy solidified at a faster rate \cite{tang_multi-length-scale_2023}. This elemental segregation is key to rationalising the hydrostatic strains; Ti additions give rise to a significant increase ($\sim$0.004\,\AA/at.\%) whilst Co, for example, seldom causes a change \cite{mishima_lattice_1985}.
Solutes that increase in lattice parameter will invoke a pseudo-hydrostatic strain; this describes the unit cell change with respect to a reference lattice parameter used for all $\gamma$ grains. 
The solute redistribution can therefore be directly inferred from the hydrostatic strain maps (Figure~\ref{fig:2}).
The Ti and Nb enrichment at the cell boundaries (as shown in Supplementary Figure 8) and their potency in changing the lattice parameter unambiguously indicates that the hydrostatic strain is induced from chemical segregation, rather than a mechanical stress.
This effect was also evident in a remelted region at a layer boundary, observed from a change in solidification direction within a $\gamma$ grain (Figure~\ref{fig:2}c), resulting in a local hydrostatic strain increase due to chemical segregation.

\subsubsection*{$\text{MC}\leftrightarrow\gamma$ orientation relationships}
The texture index of $t=1.77$ indicates a weak $\text{MC}\leftrightarrow\gamma$ OR.
This contradicts some prior Transmission Electron Microscopy (TEM) studies in Ni-based superalloys \cite{dong_microstructure_2012, xing_ni-base_2024, jiang_crystallography_2022, matysiak_microstructure_2013, markanday_microstructural_2023} that proposed strong and consistent ORs, although exceptions do occur where no OR was found \cite{lvov_mechanism_2004}.
While most of the cited studies investigated casting or powder pressing samples, where cooling rates are substantially slower than in DED-LAM, \cite{markanday_microstructural_2023} found an OR of $\hkl[110]_{\text{MC}} \parallel \hkl[110]_\gamma$ in an L-PBF-produced sample of another Ni-based superalloy, CM247LC, post-heat-treatment. 
However, recent work on the evolution of MC carbides at elevated temperatures (\SI{750}{\celsius} for \SIrange{50}{5000}{\hour}) reported a cube-cube OR that did not change with increasing exposure time at elevated temperature, despite a change in interfacial plane \cite{jiang_crystallography_2022}.
This may indicate that the effect of cooling rate and heat-treatment on the MC orientation is less significant, and that other factors, such as the MC chemical composition, could dominate the observed OR.
Additionally, TEM characterisation inherently limits the number of carbides studied due to the small sample size -- work by \cite{dong_microstructure_2012} found four observed orientation relationships in a study of 20 MC carbides, compared with over \num{37000} MC carbides explored in this work.

\subsubsection*{MC distribution}
During solidification, the growing $\gamma$ sub-grain cells reject solutes, enriching certain elements in the intercellular liquid, which ultimately promotes primary carbide formation at the cell boundaries and controls their spatial distribution.
The carbides are enriched in refractory elements Ti, Ta \& Nb \cite{tang_multi-length-scale_2023}, evident from a prior study \cite{mcauliffe_quantitative_2021} in the L-PBF condition.
In Figure~\ref{fig:3}(a), a single 2D slice of the 3D data volume is presented, where carbides clearly decorate the grain and cell boundaries in regions of locally increased strain.

The identified carbide architecture has several technological implications.
During elevated-temperature service or post-processing heat treatments, the primary MC carbides may decompose into other less stable types, via the reactions ${\rm{MC}}+ \gamma \rightarrow \rm{M_{23}C_{6}}$ or $\textrm{MC} + \gamma \rightarrow \rm{M_{6}C} + \gamma'$ \cite{reed_superalloys_2006}. 
The volume change associated with such carbide transformations can impose interfacial stresses and dislocations within the $\gamma$ matrix, creating preferential sites for microcracking \cite{song_analysis_2022}.
Factors relevant to the current study that would exacerbate the propensity for cracking include local distortions that arise from a mismatch in thermal expansion coefficients and lattice parameters of the phases due to chemical segregation.
Microcracking may occur when interfacing strengths are exceeded, either from process-induced residual stresses or when coinciding with externally applied service stresses.
In this context, the implications of local deformation patterning are next considered.

\subsubsection*{Deformation patterning}
\begin{figure}[h!]
    \centering
    \includegraphics[width=\textwidth]{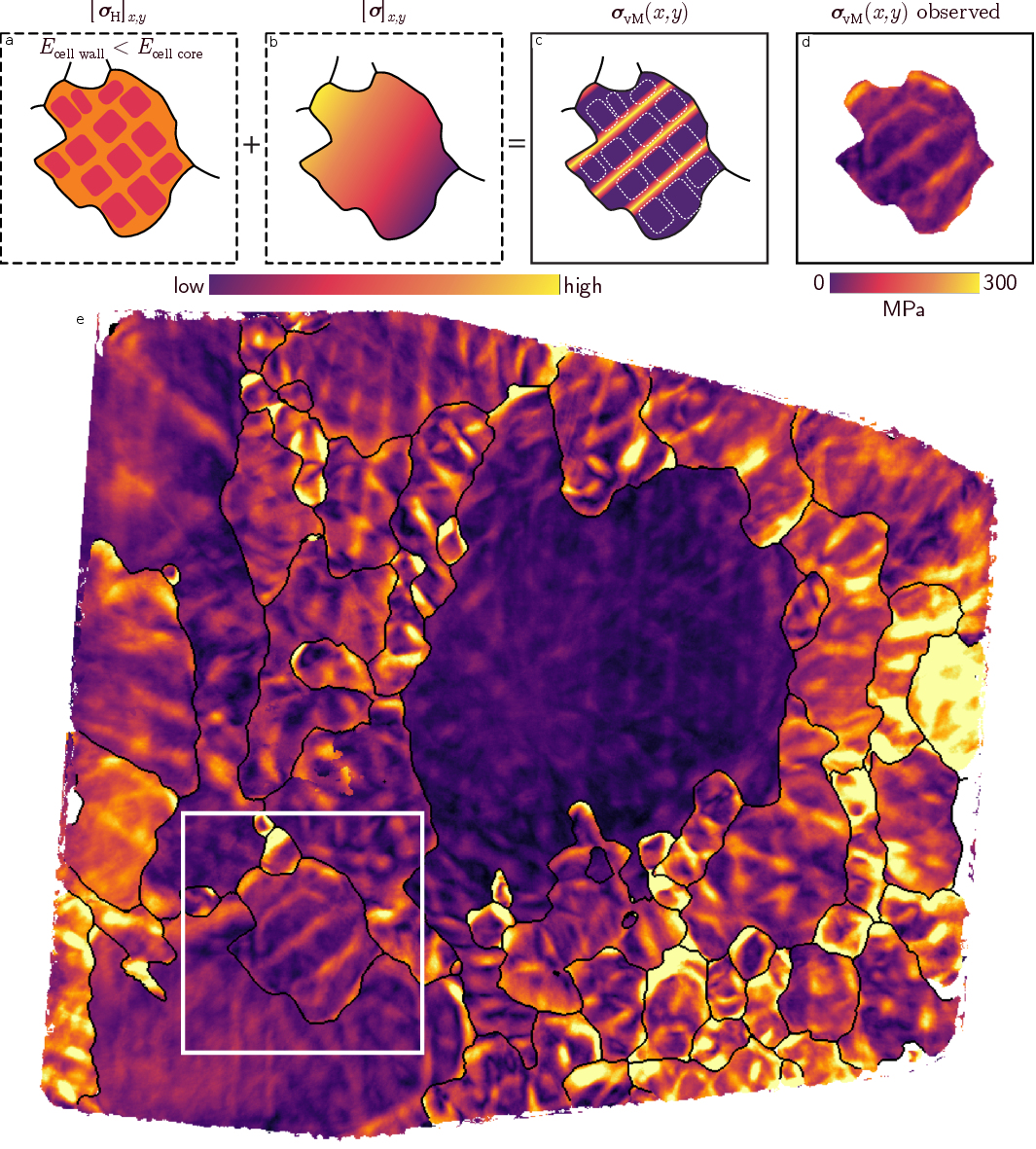}
    \caption{\textbf{Proposed residual stress formation mechanism.}  Hydrostatic stress field (a), combined with intragranular stress field (b), yields proposed von-Mises stress field (c) which is compared with observation (d), a grain-of-interest from a full 2D slice (e) of the sample volume.}
    \label{fig:4}
\end{figure}

Stress shielding is an established description that explains the redistribution of stresses within Ni-based superalloy microstructures \cite{zhang_stress_2023,griffiths_combining_2020}. 
In the case of the material studied here, stress shielding refers to the uneven sharing of load between different microstructural regions, caused by the high internal residual stresses and heterogeneous microstructure that DED-LAM introduces. 
In conventionally processed commercial alloys, the compositional and crystallographic differences between the $\gamma$ and $\gamma'$ precipitates correspond with different elastic moduli for the respective phases, where $E_{\gamma}<E_{\gamma'}$.
This effect is an important contribution to rafting during creep \cite{ohashi_elastic_1997}. 
Whilst there are no similar coherent precipitates in this alloy, the columnar microstructure with cellular-subgrain solidification pathway is analogous.

To assess the residual mechanical stresses, scalar von-Mises stresses were calculated with single-slice maps presented (Figure \ref{fig:4}(e)).
Significant stress patterning is present in many grains, with hotspots exceeding \qty{300}{\mega\pascal}.
Regions possessing higher stresses are generally located (i) near grain boundaries, or (ii) at the cell boundaries, correlating to locations where $\varepsilon_{\rm hydro} \approx 0$  (Figure \ref{fig:2}).
For (ii), it is noted that not \textit{all} cell boundaries have a high von-Mises stress; however, such high stresses tend to correlate with the cell boundaries.
A good example of this is the highlighted grain (Figure \ref{fig:2}b, Figure \ref{fig:4}a-d), where the stress forms diagonal bands that track regions where $\varepsilon_{\rm hydro} \approx 0$.
It is also unambiguous that the grains possessing the highest magnitude stresses correlate with those with the biggest range of hydrostatic strain.
For example, the large central grain has low magnitude stresses \& strains ($\sigma_{\rm vM} < \qty{100}{\mega\pascal}$ \& $\varepsilon_{\rm hydro}=\pm \num{0.2e-3}$), compared to those on the boundary exhibiting much higher magnitudes ($\sigma_{\rm vM} > \qty{300}{\mega\pascal}$ \& $\varepsilon_{\rm hydro}=\pm \num{0.5e-3}$).

The hydrostatic strain variation, arising from lattice parameter changes from chemical segregation, imposes accommodating stresses at the interface between the cell boundary and cell core due to the mismatch in respective volumes.
The morphology of the cells gives rise to complex hydrostatic stress-related, spatially varying interactions within the cell boundaries (denoted $[\boldsymbol{\sigma}]$)
The chemical composition differences between the cell boundaries and core also produce elastic moduli differences (similar to $E_{\gamma}/E_{\gamma'}$); here $E_{\rm cell\,boundary}<E_{\rm cell\,core}$.
Higher Ti concentrations, known to reduce the elastic modulus in Ni alloys \cite{kim_effects_2009}, explain the local reduction in cell boundary modulus.
These contributions are shown illustratively in Figure \ref{fig:4}(a-d).

Another factor to consider is the tortuosity of the grain structure, where the complex morphology of neighbourhood grains necessitates significant compatibility–related stresses that develop during rapid cooling.
Residual stresses operate at multiple hierarchical length scales \cite{withers_residual_2001}, and in AM alloys these comprise:
Type I (macroscale) residual stresses arise from large thermal gradients during processing. 
Type II (intergranular) stresses originate from anisotropic crystallographic orientation and grain-to-grain mechanical constraint. 
Type III (intragranular) stresses develop within cells due to segregation-induced lattice mismatch.
These stress types are well described for single-phase crystals embedded in a 3D polycrystal \cite{hayashi_intragranular_2019}.
An illustrative example of this is given in Figure \ref{fig:4}(b).

The resultant stress state in this laser additively manufactured Ni-based superalloy results from a superposition effect of (i) hydrostatic-related mismatch accommodation at cell wall/core boundaries, (ii) local $\gamma$ phase moduli variation, and (iii) intragranular stresses.
This explicitly rationalises the von-Mises stress patterning observed, as shown in Figure \ref{fig:4}(a-d).
Here, regions of high von-Mises stress are more easily accommodated by the elastically soft cell boundaries, where the directionality or modulation of these stresses must be present from the neighbourhood or compatibility-related stress contribution.
With a preference for high stresses to accumulate in the regions of high refractory element partitioning and primary MC carbides, this is undesirable.
Prior studies have often associated these cellular boundaries with sufficient angular misorientations to be classified as low-angle grain boundaries (LAGB) due to GNDs \cite{divya_microstructure_2016,chen_correlative_2021}. 
The high dislocation densities store significant energy and can initiate static recrystallisation.
The first-order estimate demonstrates the driving pressure of static recrystallisation is as high as \qty{2.9}{\mega\pascal} in L-BPF-produced ABD-900AM \cite{tang_correlative_2024}.
Given that the MC carbides are strong pinning sites for recrystallisation, their distribution at the given MC volume fraction inversely affects the pinning pressure.
Recrystallisation is therefore less restricted due to the pinning effect compared to the L-PBF process, as carbides are larger and more sparsely distributed.
In components fabricated via DED-LAM, microstructure–stress interactions must be carefully considered, especially when the as-built condition constrains structural performance. 
During fabrication, controlling thermal gradients and solidification behaviour is critical, as these factors directly influence cellular morphology/stress state and carbide distribution. 
Likewise, post-build heat treatments should be designed judiciously with the heat treatment temperatures selected for optimised microstructural features, including recrystallisation, texture, dislocation density and chemical homogeneity.

\subsubsection*{Integrating S3DXRD into the AM Characterisation Toolbox}
A range of characterisation techniques has been employed in prior studies to better understand the influence of highly non-equilibrium processing conditions on the final microstructures of AM components.
EBSD, including recent serial sectioning enhancements such as TriBeam \cite{polonsky_defects_2018}, provides key insights into crystallographic orientations and phase distributions.
The three-dimensional nature of these techniques is crucial to adequately assess spatially interconnected features resulting from solidification gradients, re-melting, and segregation that cannot be reliably inferred from 2D sections.
However, due to the inherent requirement for material removal, these techniques do not preserve the as-built strain state.
They can also be insensitive to the crystallographic orientations of particles of minor phases.
In contrast, Dark Field X-ray Microscopy (DFXM) offers non-destructive access to strain fields within individual embedded grains, improving the understanding of local intragranular processes \cite{chen_high-resolution_2023, tognan_3d_2026}.
Recent developments in pink-beam DFXM \cite{yildirim_3d4d_2025} demonstrate that multimodal approaches \cite{shukla_bridging_2025} can produce dynamic, high-resolution movies of evolving microstructures, effectively bridging the gap between mesoscale statistics and nanoscale mechanisms in situ.

Recent work by Zhang \textit{et al.} \cite{zhang_unveiling_2025}, Henningsson \textit{et al.} \cite{henningsson_microstructure_2024} and Li \textit{et al.} \cite{li_resolving_2023} has demonstrated how S3DXRD can non-destructively characterise sub-grain orientation and strain fields in large numbers of grains, addressing some limitations of both EBSD and DFXM.
Recent work by Macieira \textit{et al.} \cite{macieira_mapping_2026} has also explored the suitability of S3DXRD to investigate multi-phase materials.
Our work extends the capabilities of S3DXRD by co-registering grain morphology, sub-grain cellular networks, mechanical and chemical strains, and the spatial distribution and orientation of more than 37,000 MC carbides within a single 3D dataset.
The scanned region represents a statistically meaningful mesoscale volume, preserving build context and enabling unique insights into the coupling of residual mechanical strains and local chemical segregation.
Three capabilities enabled here are particularly significant for AM research: (i) volumetric co-registration of spatially dispersed phases with local elastic and chemical fields; (ii) preservation of the true 3D connectivity of the cellular network that mediates compatibility stresses; and (iii) the ability to identify the orientations of an extremely high number of primary and/or secondary phase precipitates.
Beyond AM alloys, this acquisition and analysis framework is a powerful tool to explore how localised chemical segregation drives 3D strain distributions and informs constitutive models in diverse multi‑phase materials.

\subsection*{Methods}

\subsubsection*{Sample preparation}
The Blown Powder Additive Manufacturing Process Replicator \cite{chen_correlative_2021} was used to perform a thin-wall bi-directional print with a laser power of \qty{200}{\watt}.
The feedstock was an argon gas-atomised ABD-900AM Ni-based superalloy powder produced by Aubert \& Duval with a mean size of \qty{30}{\micro\metre}.
The nominal composition of the alloy is Ni (bal.)-17Cr-19.9Co-3.1W-2.1Al-2.4Ti-2.1Mo-1.8Nb-1.4Ta-0.05C-0.005B \cite{tang_effect_2020}.
The thin wall print was mechanically thinned to a depth of \qty{\sim 200}{\micro\metre}.
A section near the top of the thin wall was then laser micromachined to expose an isolated cuboidal region of \qtyproduct{\approx 200 x 200 x 175}{\micro\metre\cubed} as described in Supplementary Figure 1.

\subsubsection*{Electron microscopy}
To independently verify the microstructure observations made from the S3DXRD scans, electron microscopy was performed on a cross-section of the measured sample. A Zeiss Merlin field emission gun scanning electron microscope with a secondary electron detector was used to capture the microstructure under \qty{15}{\kilo\volt} and a beam current of \qty{10}{\nano\ampere}. Prior to EDS measurement, the sample was ion polished using a Gatan PIPS system at \qty{8}{\kilo\volt} and inclined at \qty{8}{\degree}. 
EDS mapping was conducted on a dual-beam scanning electron microscope Helios 5 Hydra from ThermoFisher Scientific Inc. The electron beam was set to \qty{20}{\kilo\volt} at a beam current of \qty{3.2}{\nano\ampere}. 

\subsubsection*{S3DXRD Acquisition}
To extract highly localised crystallographic orientation and phase information in a three-dimensional volume, Scanning Three-Dimensional X-Ray Diffraction (S3DXRD) \cite{hayashi_polycrystal_2015} on the Nanoscope station \cite{wright_new_2020} of the ID11 beamline at the European Synchrotron Radiation Facility (ESRF) was employed, as shown in Figure~\ref{fig:1}(b).
A \qty{65}{\kilo\electronvolt} monochromatic X-ray beam was focused to a beam size of \qty{250}{\nano\metre}.
A S3DXRD layer was acquired via a sequence of sample rotations -- the sample was rotated \qty{360}{\degree} in $\omega$ with a step size $\delta\omega = \qty{0.125}{\degree}$ while diffraction images were collected with a DECTRIS Eiger2X CdTe 4M detector at an exposure time of \qty{0.005}{\second} per frame.
Between each \qty{360}{\degree} rotation, the diffractometer was translated along $Y$ via the $\Delta y$ motor with a step size equal to the beam size.
\num{541} rotations were collected for each S3DXRD layer, defining a slice of reconstruction space in the $XY$ plane.
\num{18} layers were collected overall, giving a total measured volume of \qtyproduct{200 x 175 x 90}{\micro\metre\cubed} with a \qty{5}{\micro\metre} $Z$ (vertical) spacing between layers.
The sample was oriented such that the build direction was approximately parallel to the laboratory $Z$ axis.

\subsubsection*{Data Analysis}
A short summary of the primary data analysis steps is provided -- the reader is referred to the Supplementary Materials for a detailed description of each analysis step.
Reconstruction of the $\gamma$ phase grains followed the general ``tomographic'' approach employed in \cite{hektor_scanning_2019} with the open-source ImageD11 3DXRD data analysis package \cite{wright_fable-3dxrdimaged11_2024}, with several modifications to improve reconstruction quality.
This generated a three-dimensional voxelised map of grain labels, where each grain had an associated grain-averaged orientation determined during the indexing stage.
This map of grain-averaged orientations was then used as the starting point for a per-voxel local orientation and strain tensor refinement process \cite{henningsson_microstructure_2024} recently incorporated in the ImageD11 3DXRD data processing package \cite{wright_fable-3dxrdimaged11_2024}.

The MC carbides, of A1 crystal structure, were indexed and mapped with a different approach.
Diffraction peaks from the $\gamma$ phase were removed, then peaks from the MC carbides were isolated via their expected $d^*$ values and intensity-filtered to remove extremely weak signals.
Peaks were paired using a Friedel-pair based approach similar to recent work by \cite{jacob_exploiting_2024}.
MC carbides were associated with their parent $\gamma$ grain if their fitted position (from the Friedel pair triangulation) overlapped with the map of $\gamma$ grain IDs determined from the tomographic reconstruction approach.
The MTEX MATLAB library \cite{hielscher_mtex_2007, bachmann_texture_2010} was used to compute misorientations between the $\gamma$ phase and MC carbides to determine pole figures.

\section*{Data Availability}
The S3DXRD experimental datasets are publicly accessible on the ESRF Data Portal under the listed DOIs: 
\href{https://doi.org/10.15151/ESRF-ES-409840930}{DOI 10.15151/ESRF-ES-409840930} \cite{chen_eco-friendly_2024-2}, \href{https://doi.org/10.15151/ESRF-ES-447646593}{DOI 10.15151/ESRF-ES-447646593} \cite{chen_eco-friendly_2024-1} and \href{https://doi.org/10.15151/ESRF-ES-473825696}{DOI 10.15151/ESRF-ES-473825696} \cite{chen_eco-friendly_2024}.
Processed per-layer S3DXRD data and associated Jupyter notebooks are available here: \href{https://doi.org/10.5281/zenodo.18619402}{DOI 10.5281/zenodo.18619402} \cite{ball_post-processed_2026}.

\section*{Code Availability}
The S3DXRD data reduction pipeline, reduced merged S3DXRD data, and figure generation scripts are available under the listed DOI: \href{https://doi.org/10.5281/zenodo.18619402}{DOI 10.5281/zenodo.18619402} \cite{ball_post-processing_2026}.

\bibliography{references}

\section*{Acknowledgments}
The authors are grateful to Carsten Detlefs, Haixing Fang, N. Axel. Henningsson, and Sir Henning F. Poulsen for their advice during the preparation of the manuscript.
C.Y. acknowledges the financial support from the ERC Starting Grant "D-REX" nr 10116911.
Y.T.T acknowledges support from the Engineering and Physical Sciences Research Council (EPSRC) for his New Investigator Award under grant number UKRI2998.
The authors are grateful for the provision and allocation of beamtime at the ESRF, under proposal number MA-4752, and for the assistance of Eleanor Lawrence Bright and Wolfgang Ludwig in collecting the experimental data.

\section*{Author Contributions Statement}
\noindent \textbf{James A. D. Ball:} Software, Formal analysis, Data Curation, Writing -- original draft, review and editing, Visualization.
\noindent \textbf{Yunhui Chen:} Conceptualization, Resources, Investigation, Methodology, Data Curation, Formal analysis, Writing -- review and editing.
\noindent \textbf{David M. Collins:} Conceptualization, Formal analysis, Investigation, Methodology, Writing -- original draft, review and editing, Visualization, Supervision.
\noindent \textbf{Yuanbo T. Tang:} Investigation, Visualization, Writing -- review and editing.
\noindent \textbf{Jonathan P. Wright:} Conceptualization, Software, Methodology, Formal analysis, Data Curation, Investigation, Writing -- review and editing.
\noindent \textbf{Can Yildirim:} Formal analysis, Writing -- original draft, review and editing.
\noindent \textbf{Julia Richter:} Investigation, Data Curation, Writing -- review and editing.

\section*{Competing Interests Statement}
The authors declare no competing interests.

\ifarXiv
    

\section*{Supplementary material}

# Revealing 3D Strain and Carbide Architectures in Additively Manufactured Ni Superalloys

James A. D. Ball[a], David M. Collins[b], Yuanbo T. Tang[c], Jonathan P. Wright[a], Can Yildirim[a], Julia Richter[d], and Yunhui Chen[e,*]

[a]European Synchrotron Radiation Facility (ESRF), 71 Avenue des Martyrs, 38000 Grenoble, France

[b]Department of Materials Science & Metallurgy, University of Cambridge, 27 Charles Babbage Road, Cambridge, CB3 0FS, United Kingdom

[c]School of Metallurgy and Materials, University of Birmingham, Edgbaston, Birmingham, B15 2TT, United Kingdom

[d]Institut für Werkstofftechnik, Universität Kassel, Mönchebergstraße 3 D-34125 Kassel, Germany

[e]Centre of Additive Manufacturing, School of Engineering, RMIT University, 124 La Trobe St, Melbourne, VIC 3000, Australia

*Corresponding author: Yunhui Chen, yunhui.chen@rmit.edu.au

# Supplementary Content

## 1 Supplementary Methods

### 1.1 Sample preparation

A diagram of the DED-LAM thin-wall printing procedure and subsequent machining is provided in Supplementary Figure 1.

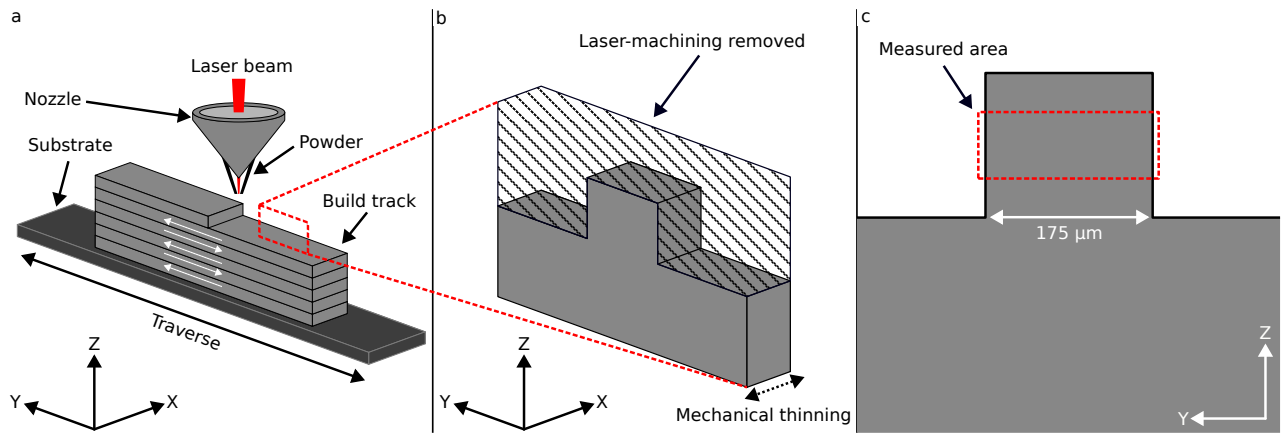

Supplementary Figure 1: (a) Thin-wall bi-directional DED-LAM print methodology. (b) Zoom on (a), showing region removed by laser machining, leaving a prominent measurable area (c).

## 1.2 S3DXRD reconstructions

Throughout this section, we refer to specific stages of the S3DXRD reconstruction pipeline via associated Jupyter notebooks and MATLAB scripts. We describe the reconstruction process in detail to clarify the choices made in handling datasets of this complexity and to provide guidance for others working on similar problems. The Jupyter notebooks and MATLAB scripts contain further procedural details, including the numerical parameters used in each step.

### 1.2.1 Software used

Supplementary Table 1 compiles the main software packages used throughout the S3DXRD data analysis pipeline in this work.

| Software | Purpose | Reference |
|---|---|---|
| `ImageD11` | Segmentation, indexing, sinogram generation | [1] |
| `Dans_Diffraction` | Structure factor determination | [2] |
| `MTEX` | Misorientation and ODF computation | [3] |
| `Nabu` | Filtered back-projection of sinograms | [4] |
| `orix` | Misorientation computation | [5] |
| `papermill` | Automated Jupyter notebook deployment | [6] |
| `Paraview` | 3D data visualisation | [7] |
| `scipy` | Friedel pair mapping | [8] |

Supplementary Table 1: Software used for S3DXRD data reduction and visualisation.

### 1.2.2 Per-layer data reduction

These data reduction tasks were performed entirely independently for each $Z$ layer, followed by the merging of layers into a contiguous dataset for further analysis (see Section 1.2.3). The `papermill` Python library [6] was employed to automate the deployment of the per-layer notebooks across all layers, ensuring consistent data analysis parameters were employed.

The notebooks produced for this study are a variant of the standard S3DXRD tomographic analysis procedure available in `ImageD11` [1] – the reader is referred to the `ImageD11` **GitHub repository** for further reference. The specific per-layer notebooks and sparse pixel files used for this study are available here: DOI 10.5281/zenodo.18619402 [9].

**Image segmentation** - `0_segment_and_label.ipynb`, `6_segment_and_label_MC.ipynb`

Segmentation of Eiger detector frames followed the standard data reduction procedure in the `ImageD11` pipeline for photon-counting detectors. A single pixel intensity threshold was chosen – pixels with intensity equal to or below the threshold were discarded. The remaining pixels was stored in a sparse array per Eiger frame containing the coordinates and intensity of each pixel. Sparse pixels were then collected across all Eiger frames into a single file (`..._sparse.h5`), then labelled to identify contiguous "2D" diffraction peaks with associated properties (such as intensity-weighted centre-of-mass positions), stored as `..._peaks_2d.h5`. These peaks were then merged across $\omega$ and $\Delta y$ to identify repeated observations of the same diffraction event on multiple frames (e.g. diffraction of a specific $(h\,k\,l)$ plane from a specific grain), producing single peaks with intensity-weighted centre-of-masses integrated over $(\omega, \Delta y)$. This process is illustrated as a "3D" merge across a single axis in Supplementary Figure 2.

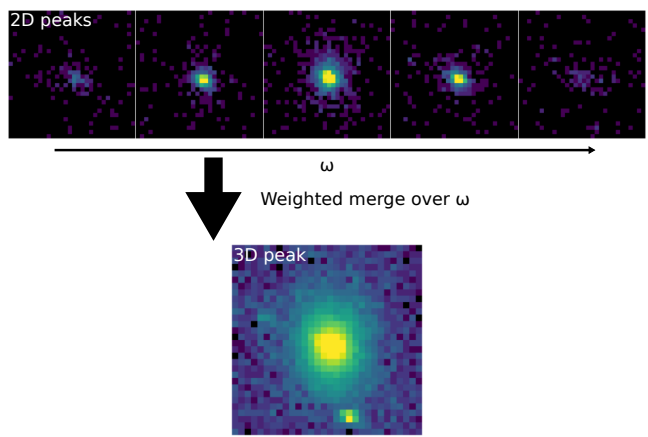

Supplementary Figure 2: 2D → 3D peak merge over $\omega$ axis.

This reduced "4D" (fast, slow, $\omega, \Delta y$) peaks dataset, stored as `..._peaks_4d.h5`, acts as an analogue to a single box-beam 3DXRD dataset where the entire layer is illuminated simultaneously, enabling the rapid identification of $\gamma$ grain orientations across the layer. During the course of the data analysis, a "beam tail" was identified, creating a weak residual diffraction signal from the large $\gamma$ grains that negatively impacted the peak merging procedure, causing disparate peaks from separate grains to be erroneously merged together into single "4D" peaks with incorrect centroids. To reduce the effect of the beam tail on the reconstruction of the $\gamma$ grains, two independent segmentations were performed with different thresholds - a higher threshold for the $\gamma$ phase to eliminate the beam tail, and a lower threshold for the (much weaker) MC carbide phase to ensure the associated peaks were not discarded. This therefore generated `..._MC` variants for the 2D and 4D peaks files.

**Phase identification** - `1_phase_identification.ipynb`

Initial determination of the phases present in the sample were performed next. Approximations of the $\gamma$ phase and MC carbide phase lattice parameters [10] were used alongside pseudo-powder pattern visualisations to approximate the observed lattice parameters required for further indexing endeavours.

**$\gamma$ phase indexing** - `2_index_Ni_simple.ipynb`

For the $\gamma$ phase, the standard `ImageD11` indexing procedure for tomographic S3DXRD data was employed, where the grain orientations were identified before being spatially mapped within the reconstruction volume. Peaks from the $\gamma$ phase were isolated via a mask on their expected $d^*$ positions, and filtered by intensity to remove a large number of weaker peaks not required for orientation identification. While the `ImageD11` indexing algorithm will not be discussed in detail here, it comprises two stages: orientation generation, where candidate orientations are predicted based on observed angles between peak scattering vectors, and orientation scoring, where each candidate orientation is scored depending on how many peaks it indexes in the dataset. The $\gamma$ grain orientations were then saved (stored in `..._grains.h5::\Ni`) if at least 10 % of the theoretically expected peaks were indexed and within the detector bounds.

**$\gamma$ phase mapping** - `3_map_Ni.ipynb`

With the orientations of $\gamma$ grains now identified, the grain shapes could be determined. Following the approach spearheaded by Hektor et al. [11], the mapping procedure proceeded as follows. To create a mask of the overall sample shape in the scanned layer, a sinogram (($\omega, \Delta y$) map) of the total diffraction signal for the layer was first produced. Each frame from the Eiger detector represents a pixel in sinogram ($\Delta y$–$\omega$) space. By integrating each Eiger frame into a single scalar measure of intensity, then mapping the result across sinogram space, a sinogram image describing total diffracted intensity can be created (Figure 3a), which, when reconstructed via an inverse radon transform approach, gives the overall shape of the sample (Figure 3b) via a real-space map of diffracted intensity.

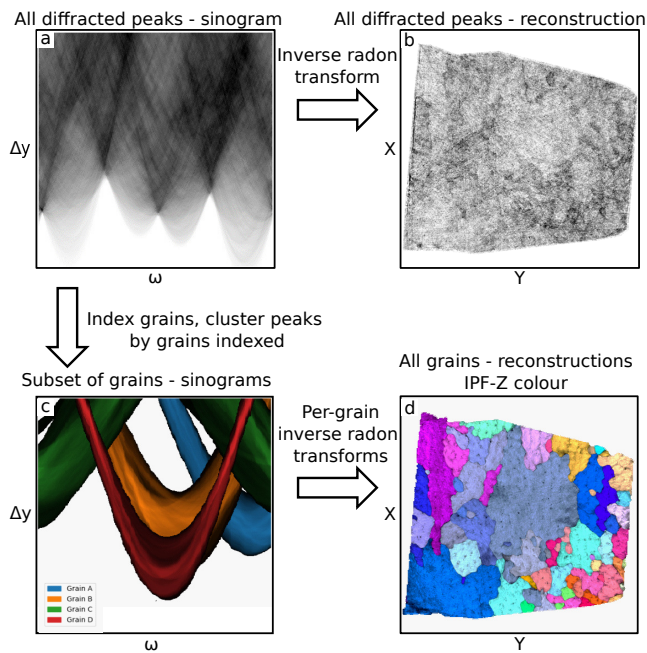

Supplementary Figure 3: Tomographic S3DXRD reconstruction process. (a) Sinogram of total diffracted signal, transformed into a real-space map of diffracted intensity (b). Per-grain sinograms (c), transformed and assembled into a real-space map of grain orientations (d).

To generate individual grain sinograms, 2D diffraction peaks were again filtered to the $\gamma$ phase and by intensity. For each $\gamma$ grain with previously-identified orientation $\boldsymbol{U}$ and reciprocal space orthogonalization matrix $\boldsymbol{B}$, 2D $\gamma$ peaks were assigned to the grain if the scattering vector in the sample reference frame, $\vec{G_s}$, of the peak resolved to near-integer $(h\,k\,l)$ values via $\vec{G}_{(h\,k\,l)} = (\boldsymbol{UB})^{-1}\,\vec{G_s}$. As this peak assignment does not account for the expected diffraction intensity of the peak, forbidden reflections were often erroneously included. To filter these before making sinogram images for each grain, the `Dans_Diffraction` Python package [2] was used to determine the expected intensity of each observed diffraction peak, thereby filtering forbidden reflections. To construct 2D sinogram images for each grain (Figure 3c), its assigned and filtered 2D diffraction peaks were binned onto $(\omega, \Delta y)$ maps, weighted by the intensity of the observed diffraction peaks after correcting for the Lorentz & polarisation factors and the structure factor for the given peak $(h\,k\,l)$, as computed by `Dans_Diffraction`.

As the raw data were collected in the half-symmetric $\Delta y$ range (from the edge to the centre of the sample, rather than edge-to-edge), tomographic reconstructions of the sinograms in this study followed a "half-acquisition" routine. For all reconstructed sinograms, an intermediate sinogram remapping was performed with the `Nabu` Python package [4] to map a $(0°, 360°)_\omega$, (-edge, 0)$_{\Delta y}$ sinogram to a geometrically-equivalent $(0°, 180°)_\omega$, (-edge, +edge)$_{\Delta y}$ sinogram which could then be processed with standard filtered back-projection inverse radon routines. Maps were originally created in $((h\,k\,l), \Delta y)$ order, as shown in Supplementary Figure 4(a), then reordered by increasing $\omega$ angle, then corrected by `Nabu` as shown in Supplementary Figure 4(b,c) $((0°, 360°)_\omega$, (-edge, 0)$_{\Delta y} \to (0°, 180°)_\omega$, (-edge, +edge)$_{\Delta y})$.

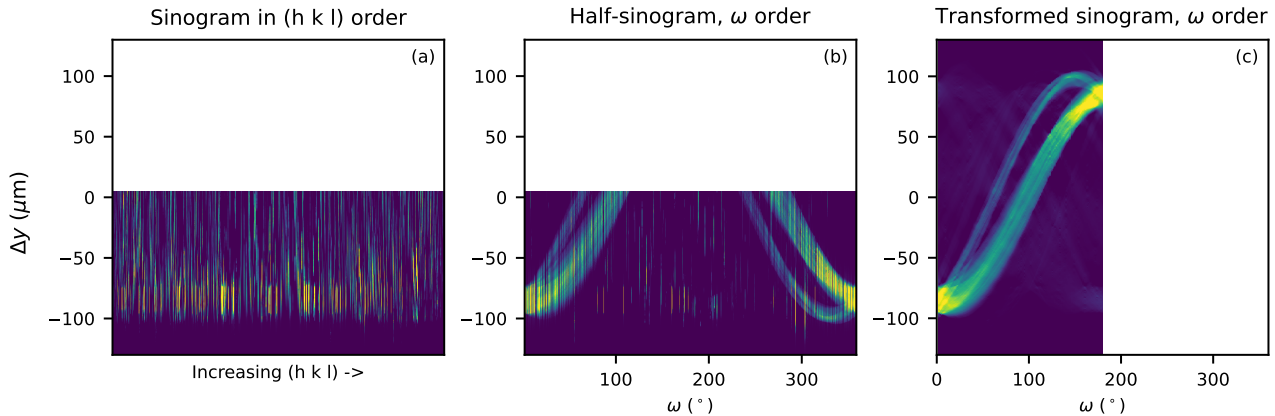

Supplementary Figure 4: (a) Sinogram for a $\gamma$ grain, in creation order $(h\,k\,l)$. (b) Sinogram reordered by increasing $\omega$ angle. (c) The same sinogram, after correcting for half-acquisition with `Nabu`.

With the sinograms generated for each grain, `Nabu` was again employed to reconstruct the shape of each grain independently via filtered back-projection. These independent reconstructions could then be combined, in a jigsaw-puzzle-style approach, into a single 2D map of grain labels per reconstruction voxel (Figure 3d), where a grain "claims" a region of the reconstruction if its shape map has the highest reconstructed intensity in that voxel. To overcome erroneous voxel assignments due to intensity fluctuations in the grain shape reconstructions (because of the MC carbides), a median filter with a kernel size of 5 was applied to the map of grain labels during this process. The associated combined diffraction intensity map was then normalised to map to a $(0, 1)$ range. The individual grain sinograms and reconstructions were exported (`..._grains.h5::\Ni\{grain id}\sinograms\ssino` and `...\recons\iradon`) alongside the combined reconstructed map (`..._grains.h5::\TensorMap_Ni`).

During the acquisition of certain layers ($z15, z20, z50, z55, z75, z80, z85$), occasional interruptions in the data were identified, either due to beam loss during the scan ($z15, z20, z55$) or encoder errors in the $\omega$ stage generating invalid motor position data. In these situations, discontinuities in the resulting grain sinograms can cause significant distortions in the reconstructed grain maps due to the filtered back-projection method employed. To improve the quality of these problematic layers, manual intervention was required to identify known-unreliable areas of the sinograms and replace them with linearly-interpolated values from the nearest reliable sinogram rows. These intervention notebooks are saved in the corresponding layer subfolders in the `PROCESSED_DATA` folder.

$\gamma$ **phase strain reconstructions** - `4_refine_Ni.ipynb`

Because $\gamma$ grains are indexed first before being mapped, the mapping notebook simply propagates the grain-averaged $\boldsymbol{U}, \boldsymbol{B}$ matrices of a given grain into the voxels assigned to that grain. The output of the previous reconstruction step can therefore be thought of as a good initial "guess" of the real map of orientation and strain state within the sample, where orientations and strains can fluctuate on a voxel-by-voxel basis. To perform a per-voxel refinement of these fields, a refinement algorithm by Henningsson *et al.* [12] has recently been integrated into `ImageD11`. The reader is referred to the cited work for a more thorough description of the reconstruction process (the supplementary material in recent work by Zhang *et al.* may also be instructive [13])– the final result of this notebook is a refined grain map, exported to

`..._refined_tmap_Ni.h5`.

**MC phase indexing and mapping** - `7_index_MC_friedel.ipynb`

With the separately-segmented peaks produced in `6_segment_and_label_MC.ipynb`, the MC carbides could be indexed. The mathematical formalism underlying the Friedel-pair-based indexing approach is now introduced. A static laboratory reference frame is defined with unit vectors $(\hat{x}_L, \hat{y}_L, \hat{z}_L)$, where the origin is determined by the intersection of the X-ray beam vector and the $\Delta y$ translation axis vector. A rotating and translating sample reference frame with unit vectors $(\hat{x}_s, \hat{y}_s, \hat{z}_s)$ is then defined, such that a transformation equation can be defined between them:

$$x_L = x_s \cos\omega - y_s \sin\omega$$
$$y_L = x_s \sin\omega + y_s \cos\omega + \Delta y - y_0 \quad (1)$$

We then define a scattering vector in the laboratory frame:

$$\overrightarrow{\Delta k_L} = \overrightarrow{k_L^{\text{out}}} - \overrightarrow{k_L^{\text{in}}} \quad (2)$$

where $\overrightarrow{k_L^{\text{in}}}$ is the incoming wave vector (along $\hat{x}_L$ in the ImageD11 formalism) and $\overrightarrow{k_L^{\text{out}}}$ is the outgoing wave vector from the origin of diffraction (assumed to be the origin of the lab frame) to the pixel position on the detector. We then rotate the scattering vector $\overrightarrow{\Delta k_L}$ by $\omega$ into the sample frame (above the rotation axis) to yield $\overrightarrow{G_s}$ (referred to in ImageD11 as a "g-vector").

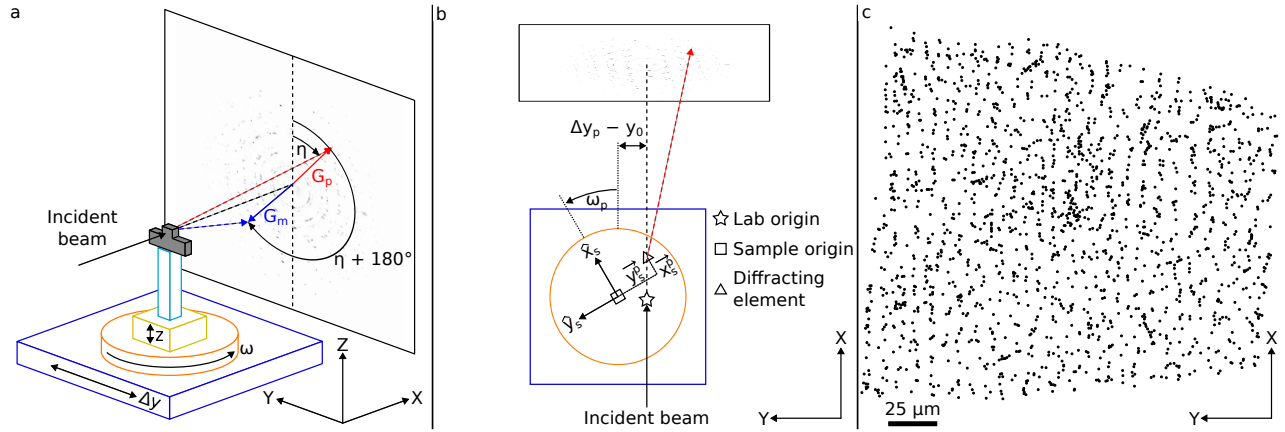

Supplementary Figure 5: (a) Diffraction geometry showing $\left(\overrightarrow{G_p}, \overrightarrow{G_m}\right)$ Friedel pairs. b) Top-down geometric construction example for $\overrightarrow{G_p}$ in the diffraction condition. (c) Resultant 2D scatter plot of identified diffraction origin positions in $(\hat{x}_s, \hat{y}_s)$ for a single scan layer.

Observed diffraction peaks can be split into "plus" and "minus" categories (the left and right sides of the detector, from the beam perspective) via the sign of their azimuthal values $\eta$. A $k$-dimensional tree can be constructed using the `scipy` Python package [8] for each side of the pair using the g-vectors $\left(\overrightarrow{G_p}, \overrightarrow{G_m}\right)$. For a valid Friedel pair, $\overrightarrow{G_p} = -\overrightarrow{G_m}$ as demonstrated in Figure 5(a). A sparse distance matrix between the two trees was constructed, yielding (plus, minus) indices to the original peaks across the identified Friedel pairs. This pairing enabled the identification of pairs of peaks that originated from the same position in the sample, but had different diffraction vectors. When a position in the sample reference frame intersects the beam and diffracts, such as the positive side of a pair as shown in Figure 5(b), the equation can be rearranged for $y_L = 0$:

$$\Delta y - y_0 = -x_s \sin\omega - y_s \cos\omega \quad (3)$$

This is calculated for given motor positions $\Delta y$ and $\omega$, where $y_0$ is the position of the $\Delta y$ motor when the rotation axis intersects the beam (nominally 0). Given a plus-minus Friedel pair $(p, m)$ where both peaks originate from the same sample position $(x_s, y_s)$ and have independent observations of $(\omega_p, \Delta y_p)$ and $(\omega_m, \Delta y_m)$ respectively, a linear system of equations can be formed and solved for the position of the origin of diffraction in the sample reference frame:

$$\begin{bmatrix} -\sin\omega_p & -\cos\omega_p \\ -\sin\omega_m & -\cos\omega_m \end{bmatrix} \begin{bmatrix} x_s \\ y_s \end{bmatrix} = \begin{bmatrix} \Delta y_p - y_0 \\ \Delta y_m - y_0 \end{bmatrix} \quad (4)$$

To apply this indexing approach, peaks from the $\gamma$ phase were filtered away, and MC peaks were selected with a small degree of intensity filtering. G-vectors were computed for each remaining MC peak, and Friedel pairs were then identified. Given the geometric formalism introduced above, the origin of diffraction in the laboratory could therefore

be triangulated for each observed Friedel pair, creating an independent spatial mapping for the diffraction signal origins without solving the crystal structure. This histogram was then segmented with ImageD11 via a connected pixels approach with a constant threshold, yielding a list of diffraction positions in the sample reference frame (Figure 5c). Then, at each diffraction position, the subset of all observed peaks within 1 μm was selected, and a local ImageD11 indexer was generated using only those selected peaks. MC carbides were retained if they indexed at least 20 diffraction peaks. Scattering vectors in the laboratory frame were re-computed before indexing, accounting for the displacement of the diffracting element along the beam. This created a list of the grain orientations found at each diffraction position, which was de-duplicated to remove repeat observations of the same grain by looking for similar grain positions and orientations, with an angular tolerance of 0.5° and a distance tolerance of 2 μm, yielding a final list of unique observed MC carbides with known positions, orientations, and strain states. The final list of de-duplicated MC carbides, with translations propagated from the positions in the 2D histogram, was then exported to `..._grains.h5::\MC`.

### 1.2.3 Post-layer analysis

The S3DXRD data reduction pipeline and reduced merged S3DXRD data are available under the listed DOI: DOI 10.5281/zenodo.18619402 [14].

$\gamma$ **grain merging** - `01_merge_Ni_layers.ipynb`, `02_mean_orientations.m`

With the layer-by-layer reconstructions fully complete, the individual layer maps could then be stacked together to form a 3D contiguous volume. While maps such as IPF-$Z$ colours and UBI matrices could be directly stacked together, identifying repeat observations of the same $\gamma$ grain across multiple slices was a more involved process. First, each layer was compared to the layer below it. $\gamma$ grains across the two layers were considered to be equivalent if their reconstructions overlapped by a minimum of 10 voxels, and if their symmetry-aware misorientation angle, computed with the `orix` Python package [5], was less than 1°. Following prior approaches (described in the Supplementary Materials: [15]), each $\gamma$ grain across all layers became a node of a graph $G$. Vertices were drawn between grains when they were determined to be equivalent. A "meta" grain could then be extracted from each connected component subgraph of $G$, describing a single $\gamma$ grain across all measured layers. To determine mean orientations for each of these "meta" grains, the MTEX MATLAB package [3] was employed to perform symmetry-aware orientation averaging (`02_mean_orientations.m`).

$\gamma \leftrightarrow$ **MC association and misorientations** - `03_carbide_misorientation.ipynb`, `04_carbide_misorientation.m`

With the $\gamma$ grains merged, the MC "child" grains were then associated with their "parent" $\gamma$ grains by taking the MC grain position, rounded to the nearest voxel coordinate, and retrieving the corresponding $\gamma$ grain ID from the 3D grain labels map. As this section of code was required in multiple subsequent notebooks for figure generation, it was moved to a utilities Python file `local_utils.py`. Then, by iterating over each "meta" $\gamma$ grain and its corresponding "child" MC grains, their orientations as Euler angles were exported to MTEX, where the misorientation matrix for each $\gamma \leftrightarrow$ MC could be determined.

Subsequent analysis was performed as required on a figure-by-figure basis – the Jupyter notebooks and MATLAB scripts for these are also provided for transparency [14].

# 2 Supplementary Results

## 2.1 Powder patterns

To identify the phases present in the DED-LAM sample, histograms of the observed diffracted intensity vs $d^*$ across all segmented diffraction peaks in all S3DXRD layers were generated, yielding a one-dimensional intensity profile; this is shown in Figure 6. Indexing of the peaks showed the microstructure comprised $\gamma$ and cubic MC carbide phases only; no evidence of the L1$_2$ structured $\gamma'$ precipitates, normally present under thermodynamic equilibrium conditions [16], was observed. Thus, the as-built DED-LAM nickel-base superalloy comprises a non-equilibrium microstructure with a highly supersaturated $\gamma$ matrix.

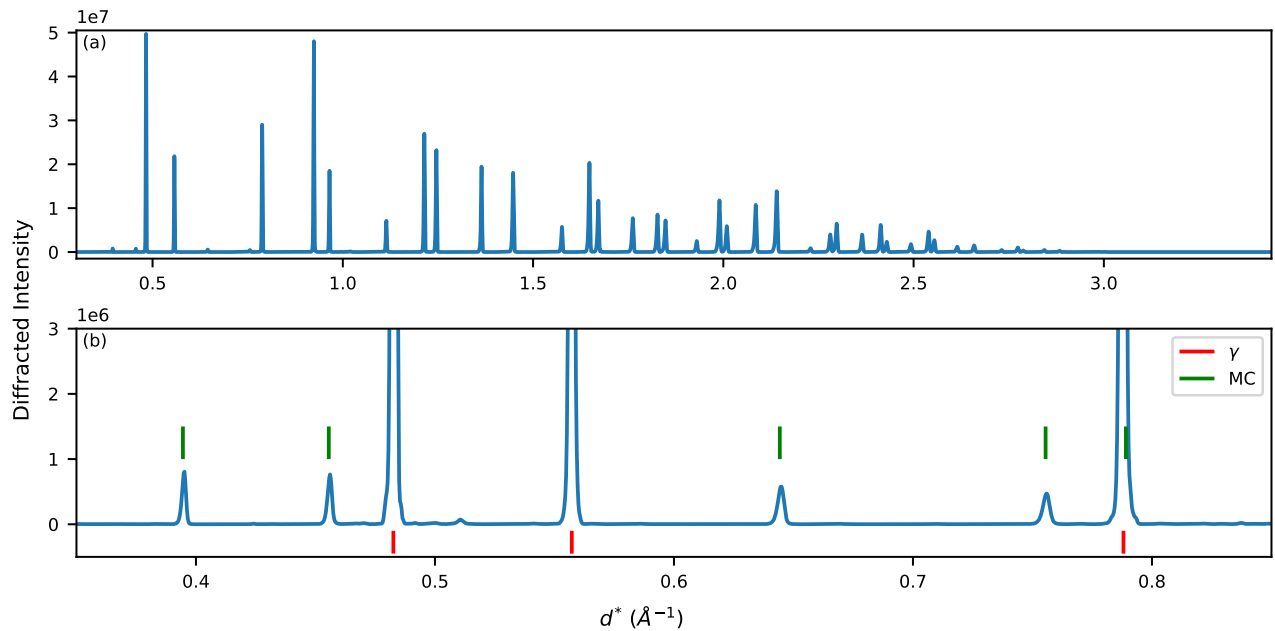

Supplementary Figure 6: (a) Integrated powder pattern across all measured layers, with zoom showing MC carbides (b)

## 2.2 EDS scans

In order to analyse the change in chemical composition, EDS mapping was performed on elements of W, Al, Co, Cr, Mo, Nb, Ni, Ta and Ti. EDS images with wt.% scale bars of each of the elements are shown in Figure 7.

An EDS line scan crossing a cell wall with a length of 6 µm was also measured as shown in Figure 8. The SEM was set to an acceleration voltage of 20 kV and a reduced beam current of 1.6 nA. A change of chemical composition was quantified with a line scan across a cell wall and cell core area. It was shown that at the sub-grain cell wall, chemical segregation occurs. The highest depletion was found for Ni, Cr and Co at the cell wall, whereas Ti and Nb are clearly increased. The elements W, Mo and Ta, however, showed a high scattering in the measured area; therefore, only tendencies were found: W seems to be slightly depleted at the cell wall, but no obvious trend in Ta or Mo concentrations is observed. This high scattering effect could originate from the higher atomic number of the elements. Therefore, the quantified results are not further interpreted.

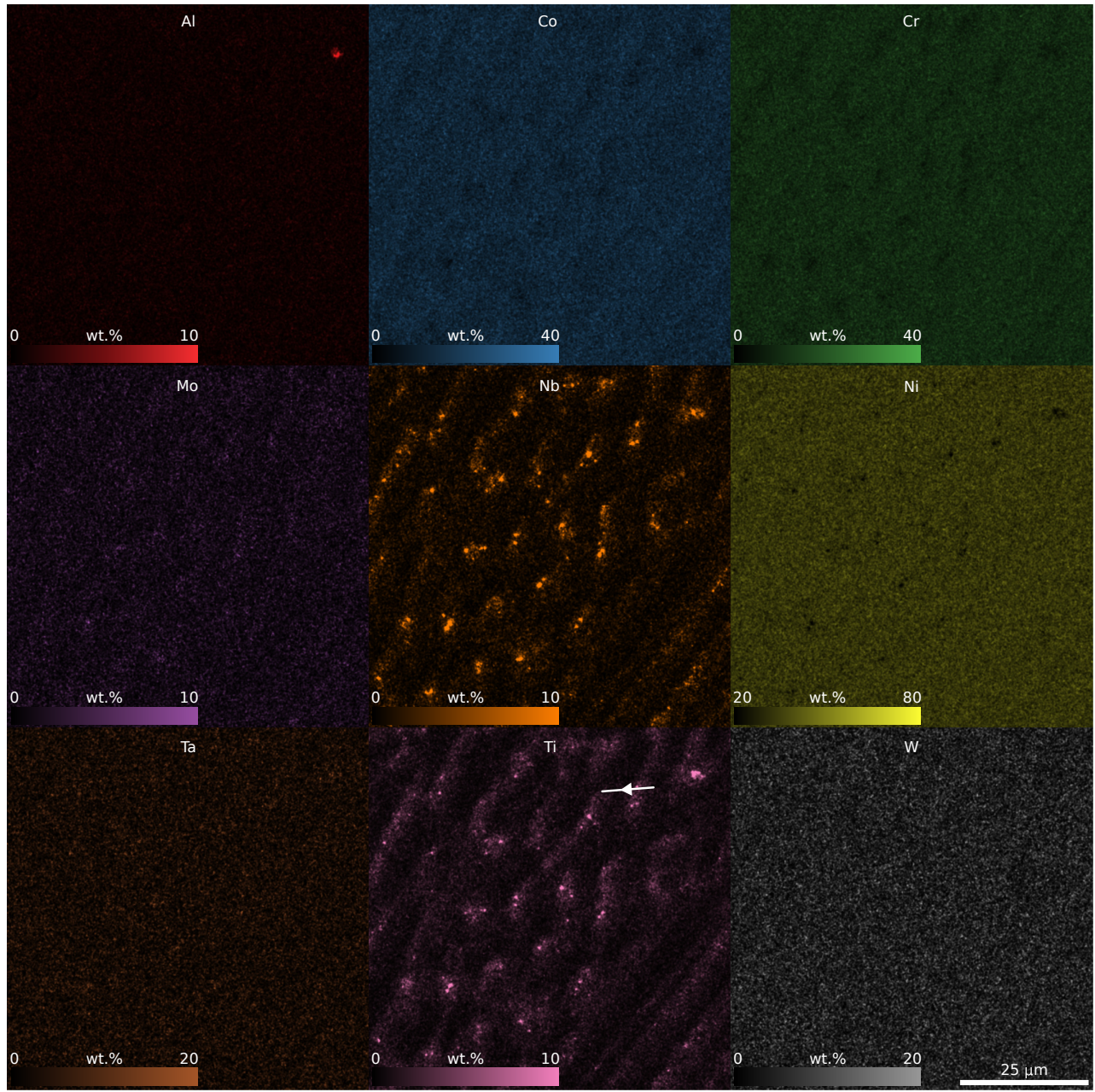

Supplementary Figure 7: EDS elemental composition maps, wt.%. White line in Ti plot illustrates approximate location of EDS line scan (Figure 8).

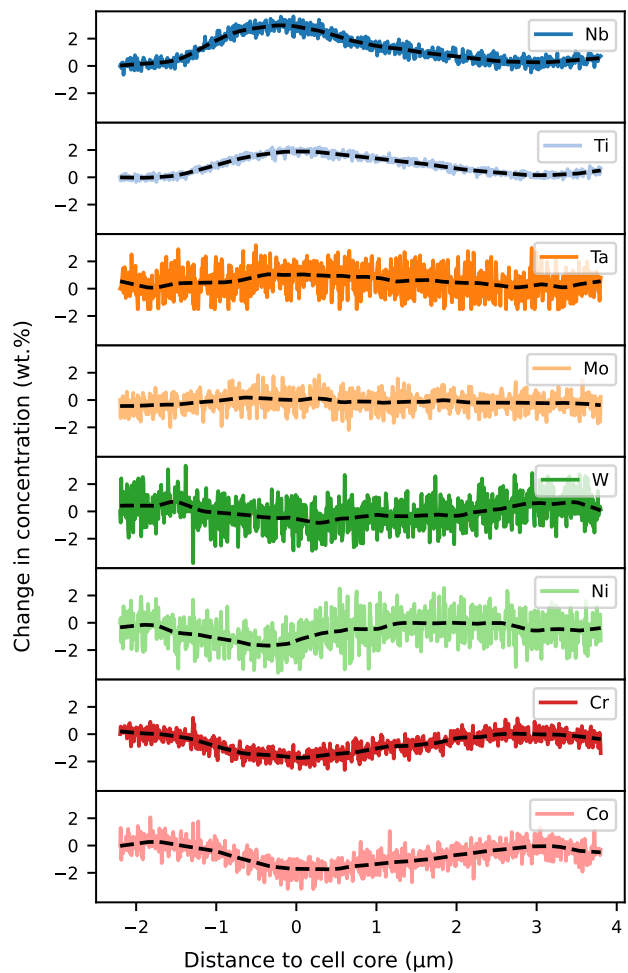

Supplementary Figure 8: ESD line scan results across a cell core, showing the relative change in composition for a range of elements. Black dashed lines represent Locally Weighted Scatterplot Smoothing (LOWESS) filters of raw data.

# Supplementary References


\fi

\end{document}